\documentclass{article}

\usepackage{lineno,hyperref}
\usepackage[a4paper, total={5.6in, 8.5in}]{geometry}

\usepackage{graphicx}
\usepackage{amsmath}
\usepackage{amsthm}
\usepackage{amssymb}
\usepackage{algorithm}
\usepackage{algorithmic}

\newtheorem{lem}{Lemma}
\newtheorem{thm}{Theorem}

\title{A Study on Load-Balanced Variants of the Bin Packing Problem}
\author{Davi Castro-Silva, Eric Gourdin}
\date{May 4, 2018}

\begin{document}
	
	\maketitle
	
	\begin{abstract}
		We consider several extensions of the fractional bin packing problem, a relaxation of the traditional bin packing problem where the objects may be split across multiple bins.
		In these extensions, we introduce load-balancing constraints imposing that the share of each object which is assigned to a same bin must be equal.
		We propose a Mixed-Integer Programming (MIP) formulation and show that the problem becomes NP-hard if we limit to at most 3 the number of bins across which each object can be split.
		
		We then consider a variant where the balanced allocations of objects to bins may be done in successive rounds; this problem was inspired by telecommunication applications, and may be used to model simple Live Streaming networks.
		We show that two rounds are always sufficient to completely assign all objects to the bins and then provide an optimal polynomial-time allocation algorithm for this problem.
	\end{abstract}

	\section{Introduction}
	
	The bin packing problem is a fundamental problem in computer science and combinatorial optimization, being a source of inspiration for the development of many approximation approaches (see, for instance, the surveys \cite{Coffman1984,CoffmanJr.2013}).
	
	In the ``classical'' version of the bin packing problem we are given a list $I$ of $n$ objects, each object $i \in I$ having a size $a_i > 0$, and a collection $J$ of empty bins each having capacity $C$ (where we assume $C \geq a_i$ for all $i \in I$).
	The goal is to pack all objects into the minimum possible number of bins, without splitting any object and without exceeding the capacity of any bin.
	
	A natural way of modeling this is by means of the following optimization problem \cite{Kantorovich1960,Martello:1990:KPA:98124}:
	\begin{eqnarray}
	\mbox{minimize}& \sum\limits_{j \in J} y_j, & \label{eq:mod1_obj}\\
	\mbox{subject to:} & \sum\limits_{i \in I} a_i x_{ij} \leq C y_j, & j \in J, \label{eq:mod1_cst1}\\
	& \sum\limits_{j \in J} x_{ij} = 1, & i \in I, \label{eq:mod1_cst2}\\
	& x_{ij}, y_j \in \{0, 1\}, & i \in I, j \in J. \label{eq:mod1_cst3}
	\end{eqnarray}
	In this description, $x_{ij}=1$ indicates that object $i$ is stored in bin $j$ and $y_j=1$ indicates that bin $j$ is used in the packing.
	Constraint (\ref{eq:mod1_cst1}) ensures that the bins' capacities are not exceeded and constraint (\ref{eq:mod1_cst2}) ensures that each object is stored in exactly one bin.
	Finally, with (\ref{eq:mod1_obj}), we are trying to minimize the total number of bins used.
	
	The bin packing problem is known to be strongly NP-hard.
	The study of its computational complexity goes all the way back to Karp's seminal paper \cite{Karp1972}, where he proved it is NP-complete to decide whether it is possible to divide a set of objects into two bins of equal size.
	
	This problem has been extensively studied and has motivated many approximation algorithms (see \cite{Coffman1984,CoffmanJr.2013}), with a considerable effort being spent trying to assess the worst-case performance ratio of simple greedy algorithms.
	For instance, in the highly influential paper \cite{Johnsonf_worst-caseperformance}, it was shown that the worst-case performance of the {\em First Fit} algorithm is less than  1.7 OPT + 1, where OPT is the optimal number of bins.
	In the same paper, the {\em First Fit Decreasing} algorithm was shown to produce a solution using at most $\frac{11}{9}$ OPT + 4 bins.
	A modified version of {\em First Fit Decreasing} was later shown to provide a guarantee of at most $\frac{71}{60}$ OPT + 1 bins being used in the packing \cite{Johnson2_worst-caseperformance}.
	More recently, the bound of $\frac{11}{9}$ OPT + $\frac{6}{9}$ bins was established for the original {\em First Fit Decreasing} algorithm and shown to be tight \cite{Dosa:2007:TBF:2399256.2399257}.
	
	The first polynomial time approximation scheme for bin packing was proposed in \cite{FernandezdelaVega1981}, where the authors proved that for any $\epsilon>0$ there exists an $O(n)$-time algorithm that gives an allocation of the objects into at most $(1+\epsilon)$OPT bins, provided OPT is sufficiently large.
	A popular approach used to obtain efficient bounds for the bin packing problem is based on the {\em Gilmore Gomory relaxation} of the set packing formulation; for instance, the procedure proposed by Karmarkar and Karp \cite{Karmarkar:1982:EAS:1382436.1382768} yields an approximate solution using at most OPT + $O(\log^2 n)$ bins.
	Several other LP models for the bin packing problem and related problems have been proposed (see \cite{ValeriodeCarvalho2002253,DelormeIoriMartello2016}).
	
	Another possible line of study is to consider continuous relaxations of the bin packing problem, obtained by relaxing the $x_{ij}$ variables in equations (\ref{eq:mod1_obj})-(\ref{eq:mod1_cst3}) to be in $[0,1]$;
	these variants represent the case where the objects may be split over several bins.
	This simple continuous version of the problem is trivial to solve (requiring only that there is enough space in the union of the bins to allocate all objects), but the problem can become interesting again when additional constraints are taken into account.
	
	In the  bin packing problem with item fragmentation \cite{10.1007/3-540-44634-6_29}, for instance, the objective is to minimize the overall cost of the packing knowing that items may be fragmented at a price.
	In a first variant the fragmentation causes an increase in the size of the fragmented items, whereas in a second variant the items' sizes do not change but the fragmentation increases their cost.
	These problems have been further studied in \cite{Shachnai2008,CASAZZA20141}, where several exact approaches and approximation algorithms are proposed.
	In a further ``cardinality constrained" variant, items can be split but each bin can only contain at most $k$ parts of items and the number of bins is to be minimized;
	approximation algorithms are proposed for this problem in \cite{Epstein2012}.
	
	In this paper, we consider another variant of the bin packing problem where items can also be fragmented, but this fragmentation must satisfy an additional load-balancing constraint.
	To the best of our knowledge, this problem is new.

	\subsection{Our problem}
	
	\smallskip
	In the {\em Balanced Fractional Bin Packing problem}, the fractions of objects assigned to a same bin should all be equal:
	if object $i$ is assigned to bin $j$, then only a proportion $\alpha_j$ of its size is stored in bin $j$ (with $\alpha_j \in \left] 0, 1\right]$ depending only on the bin $j$).
	If no additional constraints are imposed, this problem remains trivial to solve by splitting all objects equally among all available bins.
	However, on the more interesting and realistic case where we bound the maximum number of bins across which an object is allowed to be split, the problem becomes hard again:
	if each object can be split across at most $k$ bins, we prove that the problem is NP-hard when $k=1$, 2 or 3, and we conjecture it remains hard for any value of $k\in\mathbb{N}$.
	
	We then consider a multistage variant of the problem, called {\em Balanced Multistage Bin Packing problem}, where objects are assigned to bins in successive rounds or stages, a fraction of each object being assigned to (at most) a single bin at each stage.
	We also impose the balance constraint to hold for the assignments at each stage.
	
	These problems are directly inspired by the design of Live Streaming networks.
	In such networks, massive volumes of demands for videos arrive in real time at several access nodes and must be redirected towards streaming servers.
	The static design is made according to observed volumes during the peak hour, and in this design each access node is assigned to one or several servers.
	When the system is deployed, the demands arriving at each time are dispatched according to the static assignment design and the servers have to balance their streaming capabilities along the various active customers \cite{Gourdin2016}.
	
	In the design of such network delivery architectures, it is sometimes possible to assign client demands in multiple rounds, where the demands not fully satisfied during the first round are assigned to the remaining capacity of the servers in a second round.
	In practical implementations this process may be repeated several times;
	the second result conveyed in this paper shows that, in an optimal design where every client can be assigned to any server, all client demands can be fully satisfied in at most two rounds and using only the trivially minimal number of servers to accommodate them.
	
	We then provide a general polynomial-time algorithm for finding this optimal object-to-bin allocation in two rounds;
	this provides yet another example where there is a sharp barrier between problems solvable in polynomial time and NP-complete problems.

	\section{Fractional bin packing problems}
	
	A fractional bin packing problem consists in the continuous relaxation of the constraint $x_{ij}\in \{0, 1\}$ in equation (\ref{eq:mod1_cst3});
	it thus represents the case where the objects can be split over several bins.
	
	If no additional constraints are given, the minimal number of bins required will be simply $N_{frac} := \lceil S/C \rceil$, where $S := \sum_{i \in I}{a_i}$ is the sum of all objects' sizes;
	it suffices to take the objects in any order and fill the bins one at a time to obtain a solution.
	
	\subsection{The Balanced Fractional Bin Packing problem}
	\label{S:2.1}
	
	\smallskip
	Consider now a variant of the fractional bin packing problem where we require the capacity $C$ of each bin to be shared in proportion to the sizes of the objects assigned to it; in other words, if the objects in $I(j) \subset I$ are assigned to bin $j \in J$, then the bin will store the same fraction $\alpha_j$ of each object $i\in I(j)$.
	This requirement may be seen as a kind of ``fairness" constraint, useful in applications where different entities have the same right for accessing a limited number of resources.
	
	We call this problem the Balanced Fractional Bin
	Packing problem (BFBP), and it can be modeled as follows:
	\begin{eqnarray}
	\mbox{minimize}& \sum\limits_{j \in J} y_j, &\\
	\mbox{subject to:} & \sum\limits_{i \in I} a_i \lambda_{ij} \leq C y_j, & j \in J, \label{eq:cap-const}\\
	& \sum\limits_{j \in J} \lambda_{ij} = 1, & i \in I, \\
	& \lambda_{ij} = \alpha_j x_{ij}, & i \in I, j \in J, \label{eq:non-lin}\\
	& x_{ij},y_j \in \{0, 1\}, & i \in I, j \in J, \\
	& \lambda_{ij}, \alpha_j \in [0,1], & i \in I, j \in J. \label{eq:01-bound}
	\end{eqnarray}
	In these formulas, the continuous variable $\lambda_{ij}$ represents the share of the object $i$ that is stored in bin $j$; the non-linear constraint (\ref{eq:non-lin}) indicates that this share is either equal to 0 (if the object is not stored in the bin) or equal to the same value $\alpha_j$ for all objects in bin $j$.
	
	This model can be linearized in the usual fashion, replacing (\ref{eq:non-lin}) by the set of constraints:
	\begin{eqnarray}
	& \lambda_{ij} \leq x_{ij}, & i \in I, j \in J, \label{eq:lin1}\\
	& \lambda_{ij} \leq \alpha_j, & i \in I, j \in J, \label{eq:lin2}\\
	& \lambda_{ij} + 1 \geq \alpha_j + x_{ij}, & i \in I, j \in J. \label{eq:lin3}
	\end{eqnarray}
	
	Even with this additional load-balancing constraint, the fractional bin packing problem remains easy to solve:
	it suffices to take $N_{frac}$ bins of size $C$ and allocate a fraction $1/N_{frac}$ of each element into each of the bins.
	This solution, however, clearly becomes impractical when the number of bins needed $N_{frac}$ gets large.
	For a more realistic (and more interesting) variation of this problem, there should be an upper bound for the number of bins we are allowed to split each object.
	This problem will be analyzed next.

	\subsection{The $k$-Balanced Fractional Bin Packing problem}
	
	\smallskip
	In this subsection we still consider a balanced version of the fractional bin packing problem.
	However, to prevent trivial and impractical solutions such as the one presented above, we further impose an upper-limit $k$ on the number of fractions into which each object can be partitioned.
	For instance, when $k=2$ each object can be split among at most two bins, and when $k=1$ we return the classical bin packing problem.
	
	To model this problem it suffices to add the constraint:
	\begin{equation} \label{eq:k-bound}
	\sum\limits_{j \in J} x_{ij} \leq k, \hspace{5mm} i \in I.
	\end{equation}
	
	To remain consistent with the usual bin packing problem, let us make the additional assumption that each object can fit into a single bin: $\max_{i\in I} a_i \leq C$.
	An advantage of making this assumption is that the problem could only become easier to solve (by restricting the space of feasible instances), so by showing that it is already NP-hard in this case we also show that the intrinsic difficulty of the problem does not come from allowing for bigger sizes of objects.
	Moreover, the results developed in section \ref{sec:3} will always be valid in this more restricted framework, so we can compare the results of both sections over the same set of feasible instances.
	
	Formally, given an integer $k$, a set $I$ of objects whose sizes are stored in a multiset $A:=\{a_i \mbox{ }|\mbox{ } i\in I\}$, a bin size $C \geq \max_{i \in I} a_i$ and a number of bins $m$, the problem $k$-BFBP$(A, C, m)$ consists in deciding whether there exists a feasible balanced allocation of the objects to the bins where each object is split over at most $k$ bins, i.e., satisfying conditions (\ref{eq:cap-const})-(\ref{eq:01-bound}) and (\ref{eq:k-bound}) for $|J|=m$.
	
	As we have already observed, the classical bin packing problem corresponds to problem $1$-BFBP.
	As the value of $k$ grows, the constraints get more relaxed and one might wonder whether it becomes any easier to solve.
	However, this does not seem to be the case; indeed, even the particular case in which we consider the number $m$ of bins to be fixed equal to $k+1$ may be proven to be NP-complete for the first few values of $k$:
	
	\begin{thm}
		For $k\in \{1, 2, 3\}$, the decision problem $k$-BFBP$(A, C, k+1)$ on input $(A, C)$ is NP-complete.
		\label{th:1}
	\end{thm}
	
	When $k=1$, the theorem says that it is NP-complete to decide whether it is possible to divide the objects without splitting into two equal-sized bins, which was already proven by Karp when he showed that the partition problem is NP-complete \cite{Karp1972}. Theorem \ref{th:1} may then be seen as the natural generalization of this result in the case where the objects may be split over multiple bins, but satisfying a balancing (``fairness") constraint at each bin.
	
	\bigskip
	
	\noindent \underline{Proof of Theorem 1}: To prove the theorem in the case $k=2$, we will reduce every non-trivial instance of a problem already known to be NP-complete, namely the partition problem, to an instance of 2-BFBP.
	
	The partition problem is defined as follows: suppose we have a multiset $A = \{a_1, a_2, \dots, a_n\}$ of $n>1$ positive integers with sum $S:=\sum_{i = 1}^{n}{a_i}$. We then wish to know whether it is possible to partition $A$ into two subsets $A_1, A_2$ of equal sum $\frac{S}{2}$.
	
	For this partition to be possible, we must clearly have the total sum $S$ even and $\max_{i \leq n}{a_i} \leq \frac{S}{2}$, which we henceforth assume is the case (these are the non-trivial instances). Then, by denoting $r\in \{0, 1, 2\}$ the residue of $S$ modulo $3$, we prove that there exists such a balanced partition of $A$ if and only if 2-BFBP$\left( A\cup \left\lbrace \frac{S}{2}+4-r, \frac{S}{2}+4-r\right\rbrace, \frac{2(S+4-r)}{3}, 3 \right)$ is true.
	
	First note that each object can fit in a single bin, as $\max_{i \leq n}{a_i} \leq \frac{S}{2}$ by assumption and $\frac{S}{2}+4-r < \frac{2(S+4-r)}{3}$ whenever $S > 8$ (all but finitely many cases). As the sum of all elements in $A':=A\cup\left\lbrace\frac{S}{2}+4-r, \frac{S}{2}+4-r\right\rbrace$ is $2\left(S+4-r\right)$, the optimal number of used bins with size $\frac{2(S+4-r)}{3}$ must be at least 3, and in this case all bins must be completely full. But because all elements in $A'$ are integers while $\frac{2(S+4-r)}{3}$ is not, it follows that the proportionality factor of every bin must be different than 1.
	
	Each object must then be stored in exactly two bins, so if one bin has proportionality factor $\alpha \in \left]0,1\right[$ then there must be another bin with proportionality factor $1-\alpha$ (to allocate the remaining part of an object). It follows that the proportionality factors of the three bins must necessarily be $\alpha, \alpha, 1-\alpha$ for some $\alpha \in \left]0,1\right[$; w.l.o.g., let us assume $\alpha_1=\alpha_2=\alpha$, $\alpha_3=1-\alpha$.
	
	Let us now suppose $\alpha=\frac{1}{2}$, and consider all objects allocated to the first bin. Because only one half of each object is allocated to that bin and the bin is completely full with capacity $\frac{2(S+4-r)}{3}$, it follows that the sum of all these objects' sizes is $2\cdot\frac{2(S+4-r)}{3}\notin \mathbb{Z}$, a contradiction.
	
	It follows that $\alpha\neq\frac{1}{2}$, and as a result $\alpha\neq 1-\alpha$. Because of that, every element in $A'$ must have a proportion $1-\alpha$ allocated to bin 3, which implies that $\frac{2(S+4-r)}{3}=\sum_{b_i\in A'}{(1-\alpha)b_i}=(1-\alpha)\cdot2(S+4-r)$, and then $\alpha=\frac{2}{3}$.
	
	The rest is straightforward: the elements of $A'$ are distributed between bins 1 and 2 with factor $\frac{2}{3}$ each. Because $2\cdot\frac{2}{3}\left(\frac{S}{2}+4-r\right)=\frac{2(S+8-2r)}{3}>\frac{2(S+4-r)}{3}$, each object of size $\frac{S}{2}+4-r$ must be in one of these bins, which implies that there is a partition $A_1 \cup A_2$ of $A$ such that
	$$\sum_{a_i\in A_1}{\frac{2}{3}\cdot a_i}+\frac{2}{3}\left(\frac{S}{2}+4-r\right)=\frac{2(S+4-r)}{3}=\sum_{a_j\in A_2}{\frac{2}{3}\cdot a_j}+\frac{2}{3}\left(\frac{S}{2}+4-r\right)$$
	$$\Rightarrow \sum_{a_i\in A_1}{a_i}=\sum_{a_j\in A_2}{a_j}=\frac{S}{2}$$
	which is the required partition of $A$ into two subsets of equal sum.
	
	Conversely, given a partition $A_1\cup A_2$ of $A$ such that $\sum_{a_i\in A_1}{a_i}=\sum_{a_j\in A_2}{a_j}=\frac{S}{2}$, we may allocate $\frac{2}{3}$ of $\frac{S}{2}+4-r$ and of each $a_i\in A_1$ into bin 1, $\frac{2}{3}$ of $\frac{S}{2}+4-r$ and of each $a_j\in A_2$ into bin 2, and $\frac{1}{3}$ of each element in $A'$ into bin 3. This produces a solution to 2-BFBP$\left( A\cup \left\lbrace \frac{S}{2}+4-r, \frac{S}{2}+4-r\right\rbrace, \frac{2(S+4-r)}{3}, 3 \right)$.
	
	The proof for the case $k=3$ rests on similar ideas, but requires the analysis of more cases; it will be provided in Appendix \ref{ap:A}.
	\qed
	
	\medskip
	
	Although we have restricted ourselves to the case $k\leq 3$ in Theorem \ref{th:1}, we conjecture that the result is valid for all integral values of $k\geq 1$;
	the proof of this conjecture, however, will probably require much more general arguments than the ones presented here.

	\section{The Balanced Multistage Bin Packing problem}
	\label{sec:3}
	
	Given the inherent hardness of the $k$-Balanced Fractional Bin Packing problem studied in the last section, we will now consider a variation of this problem where the objects may be assigned to the bins in several stages, following a load-balancing constraint at each stage.
	
	More precisely, the assignment of the objects to the bins will be done in the following manner.
	At the first stage, a fraction of each object is assigned to a single bin in a balanced fashion.
	The proportion of an object's size not allocated to a bin at the first stage may then be served by the residual capacity of a bin at a second stage, but following the same proportionality rule for each bin;
	this process may then be repeated a certain number of times until all objects are fully allocated.
	The Balanced Multistage Bin Packing problem (BMBP) consists in finding the smallest possible number of bins of a certain capacity needed to store all objects in a fixed number of successive stages.

	\subsection{A Mixed Integer Model}
	
	\smallskip
	The model for the Balanced Fractional Bin Packing problem (section \ref{S:2.1}) can be extended to the multistage case by specifying the allowed number $k$ of successive stages, and introducing an additional index $\ell$ for these stages belonging to the set $K=\{1,\dots,k\}$, as follows:
	\medskip
	\begin{eqnarray}
	\mbox{minimize}& \sum\limits_{j \in J} y_j, &\\
	\mbox{subject to:} & \sum\limits_{i \in I} \sum\limits_{\ell \in K} a_i \lambda_{ij}^{\ell} \leq C y_j, & j \in J, \label{eq:BMBP_cst1}\\
	& \sum\limits_{j \in J} \sum\limits_{\ell \in K} \lambda_{ij}^{\ell} = 1, & i \in I, \label{eq:BMBP_cst2}\\
	& \sum\limits_{j \in J} x_{ij}^{\ell} \leq 1, & i \in I, \ell \in K, \label{eq:BMBP_cst3}\\
	& \lambda_{ij}^{\ell} \leq x_{ij}^{\ell}, & i \in I, j \in J, \ell \in K, \label{eq:BMBP_lin1}\\
	& \lambda_{ij}^{\ell} \leq \alpha_j^{\ell}, & i \in I, j \in J, \ell \in K, \label{eq:BMBP_lin2}\\
	& \lambda_{ij}^{\ell} + 1 \geq \alpha_j^{\ell} + x_{ij}^\ell, & i \in I, j \in J, \ell \in K, \label{eq:BMBP_lin3}\\
	& x_{ij}^{\ell},y_j \in \{0,1\}, & i \in I, j \in J, \ell \in K \\
	& \lambda_{ij}^{\ell}, \alpha_j^{\ell} \in [0,1], & i \in I, j \in J, \ell \in K. \label{eq:BMBP_final}
	\end{eqnarray}
	
	\medskip \noindent
	As before, the binary variable $y_j$ is equal to 1 if the $j$-th bin is used.
	The binary variable $x_{ij}^{\ell}$ is equal to 1 if object $i$ is assigned to bin $j$ at stage $\ell$ (meaning a fraction of its size is allotted to bin $j$ at this stage).
	Constraint (\ref{eq:BMBP_cst3}) ensures that at most one bin is used at each stage to allocate a given object.
	The continuous variable $\lambda_{ij}^{\ell}$ is the fraction of object $i$ allotted to bin $j$ at stage $\ell$, and all the fractions allotted to this same bin at this same stage are equal to $\alpha_j^{\ell}$ thanks to (\ref{eq:BMBP_lin1})-(\ref{eq:BMBP_lin3}) (balancing constraints).
	Constraints (\ref{eq:BMBP_cst1}) and (\ref{eq:BMBP_cst2}) are respectively the capacity and demand constraints.
	
	\smallskip
	Note that in the Balanced Multistage Bin Packing problem defined above, the number of stages $k$ is fixed and the number of bins is to be minimized.
	A closely related variant consists in fixing the number of bins (together with a choice of capacity $C$) and trying to find the minimum number of stages required to satisfy all the demands.
	More generally, being given a multiset $A$ for the objects' sizes, a bin capacity $C$, a number of bins $m$ and a number of stages $k$, problem BMBP($A, C, m, k$) consists in deciding whether there exists a feasible balanced multistage allocation of the objects to the bins, i.e., satisfying constraints (\ref{eq:BMBP_cst1})-(\ref{eq:BMBP_final}) for $|J|=m$.
	The main result of this section is a polynomial-time algorithm allowing to answer this question.
	
	\subsection{Feasibility conditions}
	\label{S:3}
	
	\smallskip
	From now on, assume the $n$ objects are sorted in decreasing order of their sizes: $A=(a_1, a_2, \dots, a_n)$ where $a_1 \geq a_2 \geq \dots \geq a_n$ (which can always be done in $O(n\log n)$ time).
	Recall that $S$ denotes the sum of all objects' sizes.
	
	Let $m$ be the minimal number of bins with which it is possible to allocate all objects in a number $k$ of successive stages according to model BMBP($A,C,m,k$).
	An immediate lower bound for $m$ is $m \geq \left \lceil S/C \right \rceil$, otherwise the total size of the objects would be bigger than the total bin capacity. 
	
	If $k=1$ (only one stage is permitted) then the problem reduces to the usual bin-packing problem, known to be NP-hard.
	%Indeed, when $k=1$ each object must be totally assigned to a single bin and the balancing constraints become redundant.
	However, if $k>1$ and the difference between any two objects' sizes is smaller than the capacity of a bin, then this lower bound can be proven to be also an upper bound:
	
	\begin{thm}
		If $a_1-a_n \leq C$ and $m:=\left \lceil S/C \right \rceil \leq n$, then it is always possible to allocate all objects into $m$ bins of capacity $C$ (which is clearly optimal) in only 2 stages.
		Moreover, such an optimal object-to-bin allocation can be obtained using an algorithm with run-time $O(n\log n)$.
		\label{th:2}
	\end{thm}
	
	Note that the condition $a_1 - a_n \leq C$ assumed by the theorem is \textit{weaker} than the assumption $\max_{i\in I}{a_i} \leq C$ which we were using before.
	This shows that, under the same restriction over the objects' sizes, problem $k$-BFBP is NP-complete (at least up to $k=3$) while problem BMBP is polytime solvable.
	
	This is especially remarkable when we note that, at any given stage, each object can be partially allocated to at most a single bin;
	so by allowing only 2 stages each object will be allocated in at most two bins, and following a proportionality constraint at each bin.
	It is not immediately clear that it should be any easier to solve this problem than the 3-BFBP, where each object is allowed to be allocated in up to 3 different bins.

	\subsection{A polynomial-time algorithm for BMBP}
	
	\smallskip
	
	To prove Theorem \ref{th:2}, we will construct an algorithm that takes as input the objects' size multiset $A$ and an integer $m$, and gives an allocation of these objects in the given number $m$ of bins of capacity $\frac{S}{m}$.
	
	This algorithm is composed of two phases: 
	\begin{itemize}
		\item{First phase:} a packing algorithm (in $n$ steps);
		\item{Second phase:} a distribution algorithm (in $m$ steps).
	\end{itemize}
	
	The first phase (Algorithm \ref{algo:2_1} below) consists in a greedy packing algorithm where the $n$ objects are grouped into $m$ ``boxes'' in order to somehow balance the amounts stored in each box.
	
	In each one of the $n$ steps (one per object), a quantity denoted $C_j$ and representing the sum of the sizes of all objects allocated to box $j \in \{1, \dots, m\}$ is updated.
	Henceforth we will call $C_j$ the ``size'' of box $j$;
	all these sizes are initially set to zero.
	
	\begin{algorithm}
		\caption{: solve BMBP (phase I) - packing algorithm}
		\begin{algorithmic}
			\STATE \underline{Initialization} ($\ell=0$):  for all $j \in \{1, \dots, m\}$, let $C_j = 0$.
			\STATE \underline{Current step $\ell$}:  the $\ell$-th object (of size $a_{\ell}$) is allocated to the box ${j}(\ell)$ such that $C_{{j}(\ell)}=\min_{1\leq j \leq m}{C_j}$, and its size is updated $C_{{j}(\ell)} \leftarrow C_{{j}(\ell)}+a_{\ell}$ (if there are more than one box with this same smallest size, we choose the one with the smallest index).
		\end{algorithmic}
		\label{algo:2_1}
	\end{algorithm}
	
	Let us denote by $\tilde{\mathcal{C}}$ the average of the box sizes after the completion of the packing algorithm:
	
	$$\tilde{\mathcal{C}} = \frac{1}{m} \sum_{j=1}^{m}{C_j} = \frac{1}{m} \sum_{i=1}^{n}{a_i}$$
	We will need the following result for the proof of Theorem \ref{th:2}:
	
	\begin{lem}
		If $a_1-a_n \leq \tilde{\mathcal{C}}$, then at the end of the packing algorithm the largest box size $C_{max}$ and the smallest box size $C_{min}$ satisfy the inequality $C_{max}-C_{min} \leq \tilde{\mathcal{C}}$.
		\label{lem:3}
	\end{lem}
	
	\noindent \underline{Proof}: Since the boxes are initially all empty, after the first $m$ steps each box contains one of the largest $m$ objects. The difference between the largest and the smallest box sizes at step $m$ is hence $a_1-a_m \leq a_1-a_n \leq \tilde{\mathcal{C}}$.
	
	Let us call $C_{max}^{(k)}$ and $C_{min}^{(k)}$ the largest and smallest box sizes at the beginning of step $k \geq m$, and proceed by induction.
	By the way the packing algorithm works we have that
	$$C_{max}^{(k+1)}=\max \left\lbrace C_{min}^{(k)}+a_k, \mbox{ } C_{max}^{(k)}\right\rbrace \mbox{ and } C_{min}^{(k+1)} \geq C_{min}^{(k)}.$$
	
	If at a given step $k > m$ the largest box size stays the same, then $$C_{max}^{(k+1)}-C_{min}^{(k+1)} = C_{max}^{(k)}-C_{min}^{(k+1)} \leq C_{max}^{(k)}-C_{min}^{(k)} \leq \tilde{\mathcal{C}}.$$
	
	Let us then suppose that at step $k > m$ the largest box size changes: $C_{max}^{(k+1)}=C_{min}^{(k)}+a_k$, and so $C_{max}^{(k+1)}-C_{min}^{(k+1)}=a_k+C_{min}^{(k)}-C_{min}^{(k+1)} \leq a_k$.
	In this case we have that
	$$m\tilde{\mathcal{C}} = \sum_{i=1}^{n}{a_i} \geq \sum_{i=1}^{k}{a_i} \geq k a_k \hspace{2mm} \Rightarrow \hspace{2mm} \tilde{\mathcal{C}} \geq \frac{k}{m}a_k > a_k \geq C_{max}^{(k+1)}-C_{min}^{(k+1)}$$
	and the lemma is proven.
	$\qed$
	
	\medskip
	During the second phase (see Algorithm \ref{algo:2_2}), the objects assigned to the boxes during the first phase are redistributed among $m$ bins in order to satisfy the balancing constraint.
	
	Define the set $X_0:=\{ C_j \mbox{ } | \mbox{ } 1 \leq j \leq m \}$ of the box sizes obtained after the packing algorithm.
	In each one of the $m$ steps (one for each box) of the distribution algorithm, the objects stored in the current box are dispatched on two bins, all having the same proportionality coefficient inside each receiving bin (corresponding to the two stages).
	
	\begin{algorithm}
		\caption{: solve BMBP (phase II) - distribution algorithm}
		\begin{algorithmic}
			
			\STATE \underline{Initialization} ($\ell = 1$):  choose $C_{\sigma(1)}:=\min\{C_j \in X_0 \mbox{ } | \mbox{ } C_j \geq \tilde{\mathcal{C}}\}$. 
			\STATE Define the quantity $S_1=C_{\sigma(1)}+\tilde{\mathcal{C}}-C_{max}$, the set $X_1=X_0\setminus \left\lbrace C_{\sigma(1)}\right\rbrace$ and the coefficients $$\lambda_1^1=\frac{S_1}{C_{\sigma(1)}}, \hspace{4mm} \lambda_m^2=1-\lambda_1^1.$$
			
			\STATE \underline{Current step $\ell$}:
			\IF{$S_{\ell-1} \leq \ell \tilde{\mathcal{C}}-C_{max}$}
			\STATE $$C_{\sigma(\ell)}:=\min \left\lbrace C_j \mbox{ } \mid \mbox{ } C_j \in X_{\ell-1} \mbox{ and } C_j \geq \tilde{\mathcal{C}} \right\rbrace;$$
			\ELSE
			\STATE $$C_{\sigma(\ell)}:=\max \left\lbrace C_j \mbox{ } \mid \mbox{ } C_j \in X_{\ell-1} \mbox{ and } C_j \leq \tilde{\mathcal{C}} \right\rbrace;$$
			\ENDIF
			\STATE Update the sum $S_\ell=S_{\ell-1}+C_{\sigma(\ell)}$ and the set $X_\ell=X_{\ell-1}\setminus \left\lbrace C_{\sigma(\ell)} \right\rbrace$.
			\STATE Define the coefficients $$\lambda_\ell^1=\frac{S_\ell-(\ell-1)\tilde{\mathcal{C}}}{C_{\sigma(\ell)}}, \hspace{4mm} \lambda_{\ell-1}^2=1-\lambda_\ell^1.$$
			
		\end{algorithmic}
		\label{algo:2_2}
	\end{algorithm}
	
	At the end of the first phase, all $n$ objects in $I:=\{1, 2, \dots, n\}$ (whose sizes are the $a_i$'s stored in $A$) have been integrally allocated to the $m$ boxes, defining the sizes $C_j$ for $j=1, \dots, m$.
	We denote by $I_1(j)$ the objects stored in box $j$ at the end of the first phase; hence
	$$I = \bigcup\limits_{j=1}^m I_1(j) \hspace{3mm} \mbox{and} \hspace{3mm} \sum\limits_{i \in I_1(j)} a_i = C_j, \hspace{3mm} j = 1, \dots, m.$$
	
	At the end of the second phase, we reassign a fraction of each object in $I$ to (at most) two bins in a circular fashion and with the aim to balance all bin sizes.
	
	More precisely, let $\lambda_{\ell}^1, \lambda_{\ell}^2$ be the coefficients and $\sigma$ be the ordering of the set $\{1, \dots, m\}$ as given by the distribution algorithm; we will use the convention $\sigma(m+1):=\sigma(1)$.
	We then assign to each bin $\ell$ ($1 \leq \ell \leq m$):
	\begin{itemize}
		\item{}On the first stage: a fraction $\lambda_\ell^1$ of each object in $I_1(\sigma(\ell))$,
		\item{}On the second stage: a fraction $\lambda_\ell^2$ of each object in $I_1(\sigma(\ell+1)).$ 
	\end{itemize}
	
	Let us illustrate the functioning of our two-phase algorithm on a small example.
	Suppose we have $n = 5$ objects of sizes 8, 7, 6, 5 and 4, and we wish to pack them into bins of capacity $C = 10$.
	For convenience we will denote each object by its size, so that $I = \{8, 7, 6, 5, 4\}$ is the set of objects to be packed.
	
	On the first phase (given by Algorithm \ref{algo:2_1}), these objects are assigned in decreasing order to $m := \lceil\frac{8+7+6+5+4}{10}\rceil = 3$ initially empty boxes, always choosing the box having the smallest size.
	At the end of this phase the objects will be allocated as shown in Figure \ref{fig:ex6} below:
	
	\begin{figure}[!htb]
		\centering
		\includegraphics[width=9cm]{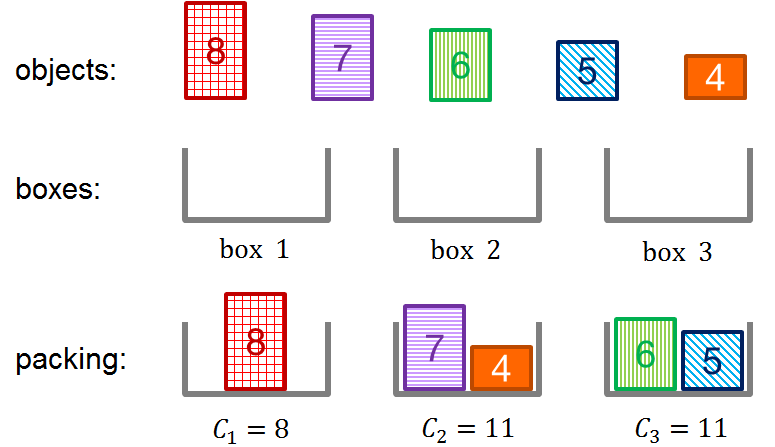}
		\caption{Packing obtained by Algorithm \ref{algo:2_1} on input $A = (8, 7, 6, 5, 4)$ and $m=3$}
		\label{fig:ex6}
	\end{figure}
	
	At the beginning of the second phase (Algorithm \ref{algo:2_2}) the objects are already assigned to the 3 boxes as above, and in our notation we have $I_1(1) = \{8\}$, $I_1(2) = \{7, 4\}$, $I_1(3) = \{6, 5\}$ and $\tilde{\mathcal{C}} = 10$.
	Following the steps of Algorithm \ref{algo:2_2}, we obtain $\sigma(1) = 2$, $\sigma(2) = 1$ and $\sigma(3) = 3$;
	the fractional allocation of the objects from the boxes to the bins proceeds as shown in Figure \ref{fig:ex7} below.
	
	\begin{figure}[!htb]
		\centering
		\includegraphics[width=11cm]{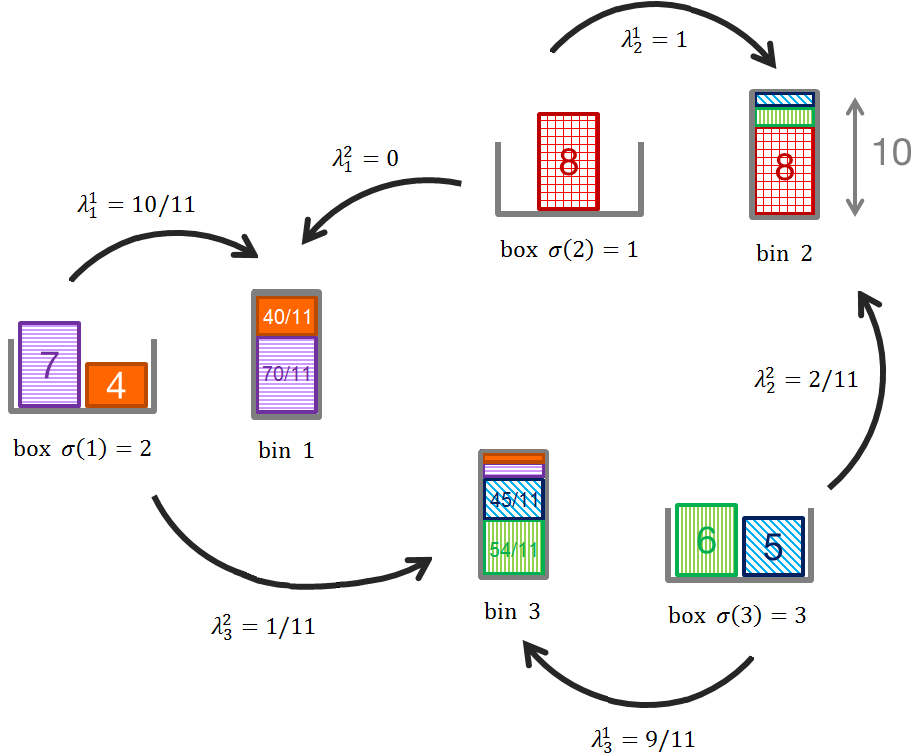}
		\caption{Balanced fractional reassignment of objects during the second phase of the algorithm.}
		\label{fig:ex7}
	\end{figure}
	
	We are now ready to prove Theorem \ref{th:2}.
	
	\subsection{Proof of Theorem \ref{th:2}}
	
	\smallskip
	
	We will show that the assignment given above satisfies all constraints imposed by the BMBP in two stages when given an instance satisfying the conditions $n \geq m := \lceil \frac{S}{C} \rceil$ and $a_1-a_n \leq \tilde{\mathcal{C}} := \frac{S}{m}$.
	If $\tilde{\mathcal{C}} < a_1-a_n \leq C$, we may increase the sizes of the smallest objects one by one until the former condition is satisfied, use the assignment described for these modified objects and then reduce them to their original sizes after the assignment is completed.
	
	First note that, by construction, at the end of the second phase all objects are entirely assigned to the bins.
	Indeed, a fraction $\lambda_\ell^1$ of the objects in $I_1(\sigma(\ell))$ is stored in bin $\ell$ and the rest ($1-\lambda_\ell^1$) is stored in bin $\ell-1$.
	
	Likewise, the total used capacity of any bin $\ell$ after the two stages is
	$$\lambda_\ell^1C_{\sigma(\ell)}+\lambda_\ell^2C_{\sigma(\ell+1)}=\left(\frac{S_\ell-(\ell-1)\tilde{\mathcal{C}}}{C_{\sigma(\ell)}}\right)C_{\sigma(\ell)}+\left(1-\frac{S_{\ell+1}-\ell\tilde{\mathcal{C}}}{C_{\sigma(\ell+1)}}\right)C_{\sigma(\ell+1)}=\tilde{\mathcal{C}},$$
	so all bins have the same used capacity $\tilde{\mathcal{C}}$.
	
	The proportionality constraint is obviously satisfied by the construction of the algorithm.
	It remains to prove only that $\lambda_\ell^1, \lambda_\ell^2 \in [0,1]$ for all $\ell \in J$, which means that a positive proportion of each object is allocated to a bin at each stage.
	
	We note that it suffices to prove that 
	$0 \leq \lambda_\ell^1 \leq \min \left \{1, \mbox{ } \tilde{\mathcal{C}}/C_{\sigma(\ell)}\right \}$ for all $\ell \in J$. Indeed:
	$$\lambda_\ell^1C_{\sigma(\ell)}\leq\tilde{\mathcal{C}} \hspace{2mm} \Rightarrow \hspace{2mm} \lambda_\ell^2=\frac{\tilde{\mathcal{C}}-\lambda_\ell^1C_{\sigma(\ell)}}{C_{\sigma(\ell+1)}} \geq 0.$$
	
	We prove this result by induction on $\ell$:
	
	\smallskip
	
	\noindent \underline{Step $\ell = 1$}: at the first step, we have $\tilde{\mathcal{C}} \leq C_{\sigma(1)} \leq C_{max}$.
	It follows that $S_1 = C_{\sigma(1)} + \tilde{\mathcal{C}} - C_{max} \leq \tilde{\mathcal{C}}$, and hence $\lambda_1^1 = S_1/C_{\sigma(1)} \leq \tilde{\mathcal{C}}/C_{\sigma(1)} \leq 1$.
	
	The result that $\lambda_1^1 \geq 0$ is a consequence of the fact that $C_{max} - C_{min} \leq \tilde{\mathcal{C}}$ (by Lemma \ref{lem:3}).
	
	\smallskip
	
	\noindent \underline{General step $\ell$}: we assume the result is true up to step $\ell$ and we'll show it still holds for step $\ell+1$.
	As in the algorithm, we distinguish two cases:
	
	\begin{enumerate}
		\item{Case when $S_\ell \leq (\ell+1)\tilde{\mathcal{C}} - C_{max}$}: in this case we have 
		$C_{\sigma(\ell+1)} \geq \tilde{\mathcal{C}}$, or equivalently $\tilde{\mathcal{C}}/C_{\sigma(\ell+1)} \leq 1$.
		It follows that 
		\begin{equation}
		\lambda_{\ell+1}^1 = \frac{S_{\ell}+C_{\sigma(\ell+1)} - \ell \tilde{\mathcal{C}}}{C_{\sigma(\ell+1)}} 
		\leq \frac{\tilde{\mathcal{C}}+C_{\sigma(\ell+1)} - C_{max}}{C_{\sigma(\ell+1)}}
		\leq \frac{\tilde{\mathcal{C}}}{C_{\sigma(\ell+1)}} \leq 1,
		\label{eq:demo1}
		\end{equation}
		and the right part of the inequality holds.
		
		Rewriting $\lambda_{\ell+1}^1$ in another way, we obtain:
		\begin{equation}
		\lambda_{\ell+1}^1 = \frac{S_{\ell} - (\ell - 1) \tilde{\mathcal{C}} + C_{\sigma(\ell+1)} - \tilde{\mathcal{C}}}{C_{\sigma(\ell+1)}} = \frac{\lambda_{\ell}^1 C_{\sigma(\ell)}+C_{\sigma(\ell+1)} - \tilde{\mathcal{C}}}{C_{\sigma(\ell+1)}}
		\label{eq:demo2}
		\end{equation}
		which yields $\lambda_{\ell+1}^1 \geq 0$ by induction ($\lambda_{\ell}^1 \geq 0$) and using 
		$C_{\sigma(\ell+1)} - \tilde{\mathcal{C}} \geq 0$.
		
		\vspace{3mm}
		
		\item{Case when $S_\ell \geq (\ell+1)\tilde{\mathcal{C}} - C_{max}$}: in this case, we have 
		$C_{\sigma(\ell+1)} \leq \tilde{\mathcal{C}}$.
		Rewriting (\ref{eq:demo2}), we obtain:
		$$
		\lambda_{\ell+1}^1 = 1 + \frac{\lambda_{\ell}^1 C_{\sigma(\ell)} - \tilde{\mathcal{C}}}{C_{\sigma(\ell+1)}} 
		$$
		By induction, we have $\lambda_{\ell}^1 C_{\sigma(\ell)} \leq \tilde{\mathcal{C}}$ and hence 
		$\lambda_{\ell+1}^1 \leq 1 \leq {\tilde{\mathcal{C}}}/{C_{\sigma(\ell+1)}}$.
		
		Finally, using the expression of $\lambda_{\ell+1}^1$ in  (\ref{eq:demo1}) and the
		lower bound on $S_\ell$ defining this second case, we obtain:
		$$
		\lambda_{\ell+1}^1 = \frac{S_{\ell}+C_{\sigma(\ell+1)} - \ell \tilde{\mathcal{C}}}{C_{\sigma(\ell+1)}}
		\geq \frac{\tilde{\mathcal{C}}+C_{\sigma(\ell+1)} - C_{max}}{C_{\sigma(\ell+1)}}.
		$$
		Using again the fact that $C_{max} - C_{min} \leq \tilde{\mathcal{C}}$ and $C_{\sigma(\ell+1)} \geq C_{min}$
		yields the result $\lambda_{\ell+1}^1 \geq 0$.
	\end{enumerate}
	
	The algorithm's run-time of $O(n\log n)$ is achieved with the use of \textit{merge-sort} to sort the multiset of demands before the packing algorithm (giving a time factor of $O(n \log n)$), the use of a \textit{min-heap} in the packing algorithm (giving a time factor of $O(n \log m)$), and a self-balancing binary tree (e.g. a \textit{red-black tree}) in the distribution algorithm (giving a time factor of $O(m \log m)$).
	\qed

	\section{Conclusions}
	
	We have investigated some special cases of fractional bin packing problems, called Balanced Fractional Bin Packing problems, where an additional load-balancing constraint is introduced.
	We showed that the problem is NP-hard when each object is allowed to be split across only up to $k=2$ or $k=3$ bins, and conjecture that the it remains NP-hard for any fixed value of $k \in \mathbb{N}$.
	
	We have also proposed a variant of this problem where the allocation of objects to bins can be achieved in successive stages, following a load-balancing constraint at each stage.
	This problem naturally occurs in certain telecommunication applications:
	when multiple entities have the same right for accessing a limited number of resources, each server is required to share its functionalities equitably between the various demands it serves;
	moreover, this resource allocation process may be allowed to take place in multiple stages until all demands are satisfied.
	
	We showed that this problem can be solved in polynomial time, and provided such an algorithm for an optimal allocation.
	This result contrasts with the hardness of the similar problem 3-BFBP, where the objects may be split in up to 3 different bins, and then provides another example of a sharp separation between NP-complete problems and problems solvable in polynomial time.
	
	Mixed-Integer Programs (MIP) were also proposed for the various problems considered, thus allowing resolution of moderate size instances with a solver.

	\appendix
	
	\section{Proof of Theorem \ref{th:1} (case $k=3$)}
	\label{ap:A}
	
	For the proof of Theorem \ref{th:1} in the case $k=3$, we will need the following lemma:
	
	\begin{lem}
		Given a multiset $A=\{a_1, a_2, \dots, a_n\}$ of positive integers with total sum $S$, the problem of deciding whether there exists a subset $I\subset A$ of sum $\frac{S}{3}$ is NP-complete.
		\label{lem:2}
	\end{lem}

	This problem is a special case of the subset sum problem, which is known to be NP-complete \cite{Martello:1990:KPA:98124}.
	It is possible to reduce every nontrivial instance of the partition problem to an instance of this restricted subset sum problem, which implies that this special case is still NP-complete.
	We omit the standard details.

	To prove Theorem \ref{th:1}, suppose we are given a multiset $A$ of positive integers with sum $S$ and we wish to decide whether there exists a subset of $A$ with sum $\frac{S}{3}$;
	by Lemma \ref{lem:2}, we know this decision problem to be NP-complete.
	We may assume without loss of generality (as we will do for the rest of the proof) that $S$ is divisible by 3 and that each element in $A$ has size at most $\frac{2S}{3}$, otherwise it is obvious that the problem cannot have a solution.
	
	We will then show how to construct from $A$ a multiset $A'$ (in linear time) with $\Theta\left(|A|\right)$ elements and sum $S'=\Theta\left(S\right)$ with the property that 3-BBP$\left(A', \frac{S'}{4}, 4\right)$ is true if and only if there exists a subset of $A$ with sum $\frac{S}{3}$; this will imply that it is NP-complete to determine whether 3-BBP$\left(A', \frac{S'}{4}, 4\right)$ is true, which is what we want to prove.
	
	Specifically, let $r\in\{0, 1\}$ be the residue of $S$ modulo 2 and define the multiset $A' :=A \cup\left\lbrace\frac{2S}{3}+r+1, \frac{2S}{3}+r+1, \frac{2S}{3}+r+1\right\rbrace$, which has sum $S':=3(S+r+1)\equiv 1 \left(\textrm{mod 2}\right)$. Consider whether the objects in $A'$ may be allocated into 4 equal bins of size $\frac{S'}{4}=\frac{3(S+r+1)}{4}$, in such a way that every object is allocated to at most 3 bins respecting a proportionality constraint at each bin; this is the problem 3-BBP$\left(A', \frac{S'}{4}, 4\right)$.
	
	Note that we are indeed in the domain of problem BBP, since each element is positive and smaller than the size $\frac{3(S+r+1)}{4}$ of each bin when $S>3$. We note also that in any solution to the problem each bin must be completely full.
	
	Suppose then that 3-BBP$\left(A', \frac{S'}{4}, 4\right)$ is true for this multiset $A'$, and consider one solution to this problem.
	Let us denote by $\alpha_{i}$ the proportionality factor of bin $i$ in this solution and by $s_{i}$ the total sum of all elements which are partially allocated to bin $i$, for $1\leq i\leq 4$. This way we must have that $s_{i}\in\mathbb{N}$ (because the elements of $A'$ are positive integers) and
	$$\alpha_{i}s_{i}=\frac{S'}{4} \hspace{2mm} \Rightarrow \hspace{2mm} \alpha_{i}=\frac{3(S+r+1)}{4s_{i}}.$$
	We then see that $\alpha_{i}\in\mathbb{Q}$ and the denominator of its irreducible fraction must be divisible by 4 (because $3(S+r+1)\equiv 1 (\textrm{mod 2}))$. This already rules out the values $\frac{1}{3}, \frac{1}{2}, \frac{2}{3}$ and 1.
	
	We will first show that there must exist a subset $I'\subset A'$ with sum $\frac{S'}{3}$. There are two possible cases that we must analyze depending on the values of the proportional factors $\alpha_{i}$ of the bins:
	
	\medskip
	\noindent
	\textbf{First case:} There are two of the proportional factors which sum to 1.
	
	\noindent
	By symmetry we may suppose that $\alpha_1+\alpha_2=1$, so that $\alpha_1=\beta$ and $\alpha_2=1-\beta$ for some $\beta \notin \left\lbrace \frac{1}{2}, \frac{2}{3}\right\rbrace$. We may also suppose there are at least as many objects allocated to bin 1 as there are objects allocated to bin 2.
	There are three possibilities to consider in this case:
	
	\begin{itemize}
		\item \textit{Every object allocated to bin 1 is also allocated to bin 2:}
		
		Then the whole size of these objects is allocated to the union of the two bins and must have total sum $s_1=s_2=2\cdot\frac{S'}{4}\notin \mathbb{Z}$, an absurd.
		
		\item \textit{There is an object allocated to bin 1 which is also allocated to bin 3 but not to bin 4 (or, symmetrically, to bin 4 but not to bin 3):}
		
		Then $\alpha_1+\alpha_3=1$, so that $\alpha_3=\alpha_2=1-\beta$. There are only three possible values for $\alpha_4$ in order to satisfy all constraints:
		
		\begin{enumerate}
			\item $\alpha_4+\alpha_1=1 \hspace{2mm} \Rightarrow \hspace{2mm} \alpha_4=1-\beta$:
			
			Note that $\alpha_2+\alpha_3 \neq 1$ (otherwise $\beta=\frac{1}{2}$) and $\alpha_2+\alpha_3+\alpha_4 \neq 1$ (otherwise $\beta=\frac{2}{3}$); thus all objects must be partially allocated to bin 1, implying $s_1=S'$ and $\beta=\frac{1}{4}$. This in turn implies that $\alpha_2 s_2=\frac{3}{4} s_2=\frac{S'}{4}$, so the set $I'$ of all objects partially allocated to bin 2 must have sum $\frac{S'}{3}$.
			
			\item $\alpha_4+\alpha_2=\alpha_4+\alpha_3=1 \hspace{2mm} \Rightarrow \hspace{2mm} \alpha_4=\beta$:
			
			Every object will be allocated to either bin 1 or to bin 4, but not both (this is the only way to distribute its entire size to the bins respecting the constraints). This implies $s_1+s_4=S'$, and because $\alpha_1=\alpha_4=\beta$, we have $s_1=s_4=\frac{S'}{2}\notin\mathbb{Z}$ which cannot be.
			
			\item $\alpha_4+\alpha_2+\alpha_3=1 \hspace{2mm} \Rightarrow \hspace{2mm} \alpha_4=2\beta-1$:
			
			Again every object will be allocated to either bin 1 or to bin 4, but not both. This implies that $\beta s_1=\frac{S'}{4}=(2\beta-1)(S'-s_1)$, which admits no rational solution for $\beta$ when $S'>0$, absurd.
			
		\end{enumerate}
		
		\item \textit{Every object allocated to bin 1 but not to bin 2 must be allocated to both bins 3 and 4:}
		
		This way $\alpha_1+\alpha_3+\alpha_4=1$, and because $\alpha_2\neq\alpha_1$ (otherwise $\beta=\frac{1}{2}$) we must have $\alpha_2+\alpha_3+\alpha_4\neq1$. There remains two possibilities:
		
		\begin{enumerate}
			\item \textit{Every object is partially allocated to bin 1:}
			
			Then $s_1=S'$ implies $\alpha_1=\frac{1}{4}$ and $\alpha_2=\frac{3}{4}$, so that the set $I'$ of all objects allocated to bin 2 has sum $s_2=\frac{S'}{3}$.
			
			\item \textit{There is an object not allocated to bin 1:}
			
			Then this object must be completely allocated to bins 2 and 3 (or, symmetrically, bins 2 and 4), so that $\alpha_2+\alpha_3=1 \Rightarrow \alpha_3=\beta$, $\alpha_4=1-2\beta$; in this case every object must be allocated either to bin 2 or to bin 4, but not both. This implies that $(1-\beta) s_2=\frac{S'}{4}=(1-2\beta)(S'-s_2)$ which has no rational solution for $\beta$ when $S'>0$, absurd.
		\end{enumerate}
		
	\end{itemize}
	
	%\item \textbf{Second case:} No two of the proportional factors sum to 1.
	\noindent
	\textbf{Second case:} No two of the proportional factors sum to 1.
	
	\noindent
	By symmetry, we may suppose that $\alpha_1+\alpha_2+\alpha_3=1$. The fourth factor must then be equal to one of the first three (so that it can sum to 1 when added to other two of them), so we suppose $\alpha_4=\alpha_1$. There are three possibilities to consider:
	
	\begin{itemize}
		\item $\alpha_1=\alpha_2=\alpha_3$
		
		Together with $\alpha_1+\alpha_2+\alpha_3=1$ this implies $\alpha_1=\frac{1}{3}$, impossible.
		
		\item $\alpha_1\neq\alpha_2$ and $\alpha_1\neq\alpha_3$
		
		Together with $\alpha_1=\alpha_4$ this implies that every object must be either in bin 1 or bin 4, but not both, which gives $\alpha_1=\frac{1}{2}$, impossible.
		
		\item $\alpha_1=\alpha_2\neq\alpha_3$ (or $\alpha_1=\alpha_3\neq\alpha_2$)
		
		Every object must then be partially allocated to bin 3, which implies that $\alpha_3=\frac{1}{4}$ and $\alpha_1=\frac{3}{8}$; the set $I'$ of all objects \textit{not} allocated to bin 1 must then have sum $\frac{S'}{3}$.
		
	\end{itemize}
	
	%\end{itemize}
	
	We have then shown that for 3-BBP$\left(A', \frac{S'}{4}, 4\right)$ to be true there must be a subset $I'\subset A'$ with sum $\frac{S'}{3}$. Now we note that the multiset $I'$ cannot have more than one of the elements $\frac{2S}{3}+r+1$, since $2\cdot\left(\frac{2S}{3}+r+1\right)>S+r+1=\frac{S'}{3}$, and likewise $A'\setminus I'$ can't have more than two of the elements $\frac{2S}{3}+r+1$ since $3\cdot\left(\frac{2S}{3}+r+1\right)>2S+2r+2=\frac{2S'}{3}$; this implies that there is exactly one of the elements $\frac{2S}{3}+r+1$ in $I'$, so that $I:=I'\setminus\left\lbrace\frac{2S}{3}+r+1\right\rbrace$ is a subset of $A$ with sum $\frac{S'}{3}-\left(\frac{2S}{3}+r+1\right)=\frac{S}{3}$, as we wanted.
	
	Conversely, if $I$ is a subset of $A$ with sum $\frac{S}{3}$, then we may allocate $\frac{1}{4}$ of each element in $A'$ to bin 1, $\frac{3}{4}$ of all elements in $I\cup\left\lbrace\frac{2S}{3}+r+1\right\rbrace$ to bin 2, and $\frac{3}{8}$ of all elements in $(A\setminus I)\cup\left\lbrace\frac{2S}{3}+r+1, \frac{2S}{3}+r+1\right\rbrace$ to each of bins 3 and 4, thereby satisfying all constraints of the problem.
	\qed

	\bibliography{BinPacking}

\begin{thebibliography}{10}

\bibitem{Coffman1984}
E.~G. Coffman, M.~R. Garey, and D.~S. Johnson, {\em Approximation Algorithms
  for Bin-Packing --- An Updated Survey}, pp.~49--106.
\newblock Vienna: Springer Vienna, 1984.

\bibitem{CoffmanJr.2013}
E.~G. Coffman~Jr., J.~Csirik, G.~Galambos, S.~Martello, and D.~Vigo, {\em Bin
  Packing Approximation Algorithms: Survey and Classification}, pp.~455--531.
\newblock New York, NY: Springer New York, 2013.

\bibitem{Kantorovich1960}
L.~V. Kantorovich, ``Mathematical methods of organizing and planning
  production,'' {\em Management Science}, vol.~6, pp.~363 -- 422, 1960.

\bibitem{Martello:1990:KPA:98124}
S.~Martello and P.~Toth, {\em Knapsack Problems: Algorithms and Computer
  Implementations}.
\newblock New York, NY, USA: John Wiley \& Sons, Inc., 1990.

\bibitem{Karp1972}
R.~M. Karp, {\em Reducibility among Combinatorial Problems}, pp.~85--103.
\newblock Boston, MA: Springer US, 1972.

\bibitem{Johnsonf_worst-caseperformance}
D.~S. Johnson, A.~Demers, J.~D. Ullman, M.~R. Garey, and R.~L. Graham,
  ``Worst-case performance bounds for simple one-dimensional packing
  algorithms,'' {\em SIAM Journal on Computing}, vol.~3, no.~4, pp.~229--325,
  1974.

\bibitem{Johnson2_worst-caseperformance}
M.~R. Garey and D.~S. Johnson, ``A 71/60 theorem for bin packing,'' {\em
  Journal of Complexity}, vol.~1, pp.~65--106, 1985.

\bibitem{Dosa:2007:TBF:2399256.2399257}
G.~D\'{o}sa, ``The tight bound of first fit decreasing bin-packing algorithm is
  ffd(i) $\leq$ 11/9opt(i) + 6/9,'' in {\em Proceedings of the First
  International Conference on Combinatorics, Algorithms, Probabilistic and
  Experimental Methodologies}, ESCAPE'07, (Berlin, Heidelberg), pp.~1--11,
  Springer-Verlag, 2007.

\bibitem{FernandezdelaVega1981}
W.~Fernandez de~la Vega and G.~S. Lueker, ``Bin packing can be solved within 1
  + $\epsilon$ in linear time,'' {\em Combinatorica}, vol.~1, no.~4,
  pp.~349--355, 1981.

\bibitem{Karmarkar:1982:EAS:1382436.1382768}
N.~Karmarkar and R.~M. Karp, ``An efficient approximation scheme for the
  one-dimensional bin-packing problem,'' in {\em Proceedings of the 23rd Annual
  Symposium on Foundations of Computer Science}, SFCS '82, (Washington, DC,
  USA), pp.~312--320, IEEE Computer Society, 1982.

\bibitem{ValeriodeCarvalho2002253}
J.~M. {Val\'erio de Carvalho}, ``{LP} models for bin packing and cutting stock
  problems,'' {\em European Journal of Operational Research}, vol.~141, no.~2,
  pp.~253 -- 273, 2002.

\bibitem{DelormeIoriMartello2016}
S.~Delorme, M.~Iori, and S.~Martello, ``Bin packing and cutting stock problems:
  Mathematical models and exact algorithms,'' {\em European Journal of
  Operational Research}, vol.~255, pp.~1--20, 2016.

\bibitem{10.1007/3-540-44634-6_29}
N.~Menakerman and R.~Rom, ``Bin packing with item fragmentation,'' in {\em
  Algorithms and Data Structures} (F.~Dehne, J.-R. Sack, and R.~Tamassia,
  eds.), (Berlin, Heidelberg), pp.~313--324, Springer Berlin Heidelberg, 2001.

\bibitem{Shachnai2008}
H.~Shachnai, T.~Tamir, and O.~Yehezkely, ``Approximation schemes for packing
  with item fragmentation,'' {\em Theory of Computing Systems}, vol.~43,
  pp.~81--98, Jul 2008.

\bibitem{CASAZZA20141}
M.~Casazza and A.~Ceselli, ``Mathematical programming algorithms for bin
  packing problems with item fragmentation,'' {\em Computers \& Operations
  Research}, vol.~46, pp.~1 -- 11, 2014.

\bibitem{Epstein2012}
L.~Epstein, A.~Levin, and R.~van Stee, ``Approximation schemes for packing
  splittable items with cardinality constraints,'' {\em Algorithmica}, vol.~62,
  pp.~102--129, Feb 2012.

\bibitem{Gourdin2016}
E.~Gourdin, ``Conception d'un reseau de diffusion live de videos, avec prise en
  compte de pannes de serveurs et equilibrage de charge,'' {\em ROADEF
  conference}, 2016.

\end{thebibliography}
	\bibliographystyle{ieeetr}
	
\end{document}